\documentclass[pra,twocolumn,superscriptaddress]{revtex4-1}

\usepackage[utf8]{inputenc}
\usepackage[english]{babel}
\usepackage[T1]{fontenc}

\usepackage{microtype}

\usepackage{graphicx}

\usepackage{amsmath}
\usepackage{amsfonts}
\usepackage{amssymb}

\usepackage{amsthm}
\usepackage{dsfont}

\usepackage[breaklinks=true,colorlinks]{hyperref}

\newcommand{\cf}{cf.\ }
\newcommand{\coloneq}{\mathrel{\mathop:}=}
\newcommand{\eqcolon}{=\mathrel{\mathop:}}
\newcommand{\dd}{\mathrm{d}}
\newcommand{\Tr}{\operatorname{Tr}}
\newcommand{\ew}[1]{\left\langle{#1}\right\rangle}
\newcommand{\kB}{k_\mathrm{B}}
\newcommand{\Treal}{T_\mathrm{real}}
\newcommand{\nbar}{\bar{n}}
\newcommand{\Thetaf}{\Theta_\mathrm{fic}}
\newcommand{\UB}{U_\mathrm{B}}
\newcommand{\Uwf}{U_\mathrm{WF}}

\begin{document}

\title{On the operation of machines powered by quantum non-thermal baths}

\author{Wolfgang Niedenzu}
\affiliation{Department of Chemical Physics, Weizmann Institute of Science, Rehovot~7610001, Israel}

\author{David Gelbwaser-Klimovsky}
\affiliation{Department of Chemistry and Chemical Biology, Harvard University, Cambridge, MA~02138, USA}

\author{Abraham~G.~Kofman}
\affiliation{Department of Chemical Physics, Weizmann Institute of Science, Rehovot~7610001, Israel}
\affiliation{CEMS, RIKEN, Saitama, 351-0198, Japan}

\author{Gershon Kurizki}
\affiliation{Department of Chemical Physics, Weizmann Institute of Science, Rehovot~7610001, Israel}

\begin{abstract}
  Diverse models of engines energised by quantum-coherent, hence non-thermal, baths allow the engine efficiency to transgress the standard thermodynamic Carnot bound. These transgressions call for an elucidation of the underlying mechanisms. Here we show that non-thermal baths may impart not only heat, but also mechanical work to a machine. The Carnot bound is inapplicable to such a hybrid machine. Intriguingly, it may exhibit dual action, concurrently as engine and refrigerator, with up to 100\% efficiency. We conclude that even though a machine powered by a quantum bath may exhibit an unconventional performance, it still abides by the traditional principles of thermodynamics.
\end{abstract}

\maketitle

\section{Introduction}

Scully and co-workers~\cite{scully2003extracting,scully2002extracting,rostovtsev2003extracting} have introduced a model of a Carnot heat engine based on a bath comprised of partly-coherent three-level atoms (nicknamed ``phaseonium'') that interact with a cavity-mode ``working fluid'' (WF) while they cross the cavity. Their astounding conclusion was that the efficiency of such an engine may exceed the universal Carnot bound~\cite{schwablbook} because the phaseonium bath endows the cavity mode with a temperature $T_\varphi$ that, depending on the phase $\varphi$ of the atomic coherence, may surpass the corresponding temperature of thermal atoms without coherence. This research has initiated diverse suggestions of quantum resources for boosting heat-machine efficiency above the Carnot bound, with a focus on non-thermal baths possessing quantum coherence~\cite{dillenschneider2009energetics,deliberato2011carnot,huang2012effects,abah2014efficiency,li2014quantum,rossnagel2014nanoscale,hardal2015superradiant,turkpence2016quantum}.

\par

In traditional heat engines, energy exchange between the WF and the (hot and cold) thermal baths are entropy-changing heat transfers, whereas parametric changes of the WF Hamiltonian are isentropic processes that produce or invest work~\cite{alicki1979quantum,geva1992quantum}. The main questions we raise here are: Does the same division between heat and work necessarily hold in engines fuelled by non-thermal (quantum-coherent) baths and how does this division affect the engine efficiency? To what extent is the quantum character of non-thermal baths and (or) the WF relevant to the engine performance?

\par

Here we address the above questions by means of the fundamental notion of \emph{non-passivity}~\cite{pusz1978passive,lenard1978thermodynamical} that defines the ability of a quantum state to deliver work. The maximal work extractable from a non-passive state is known as ergotropy~\cite{allahverdyan2004maximal,binder2015quantum} (\ref{app_ergotropy}). The significance of non-passivity as a work resource has been previously demonstrated for a heat machine with a quantised piston~\cite{gelbwaser2013work,gelbwaser2014heat}.

\par

By resorting to this notion, we point out that there are two kinds of machines fuelled by non-thermal baths. In machines of the first kind (exemplified by the intriguing~\cite{abah2014efficiency,rossnagel2014nanoscale}) the energy imparted by a non-thermal bath to the WF consists of an isentropic part that transfers ergotropy (work) to the WF, which has hitherto been unaccounted for, and an entropy-changing part that corresponds to heat transfer, but the total energy received by the WF cannot be associated with heat. By contrast, in machines of the second kind (exemplified by the pioneering~\cite{scully2003extracting,scully2002extracting,rostovtsev2003extracting}) the entire energy transfer from the non-thermal bath to the WF can indeed be considered as heat. A correct division of the energy transfer from the bath to the WF into heat and work is crucial for the realisation that the efficiency of machines of the first kind does not have a thermodynamic bound that may be deduced from the second law. This becomes evident when the energy of the non-thermal bath has a vanishingly small thermal component: The engine can then produce work without heat input.

\par

Our analysis of these two kinds of machines is focused on an Otto cycle for an harmonic-oscillator WF under the assumption that the non-thermal bath that powers the machine is unitarily generated from a thermal one. A central result of this analysis is that such non-thermal baths may produce a non-passive steady state of the WF and thereby change its ergotropy. We use this result to identify the two distinct kinds of machines powered by quantum non-thermal baths:

\par

(i)~Machines of the first kind are exemplified by setups fuelled by a squeezed thermal bath or a coherently-displaced thermal bath~\cite{kim1989properties,oz1991thermal,marian1993squeezed}) which render the WF state \emph{non-passive} (and therefore non-thermal). Our central finding is that this kind of machine does not act as a heat engine, but rather as a \emph{hybrid thermo-mechanical machine energised by work as well as heat} imparted by this bath. The thermodynamic Carnot bound does not apply to the efficiency of such a machine, which is shown to operate not only as an engine, but concurrently as a heat pump/refrigerator that moves heat from the ``cold'' bath to the ``hot'' non-thermal bath, at the expense of mechanical work invested by the latter.

\par

(ii)~Machines of the second kind are obtained for WF--bath interactions whereby, in contrast to machines of the first kind, the WF is rendered \emph{thermal} (i.e., passive) by the non-thermal bath. An engine fuelled by a phaseonium bath~\cite{scully2003extracting,scully2002extracting,rostovtsev2003extracting,turkpence2016quantum} exemplifies this kind of machines. It is shown to act as a genuine heat engine, whose efficiency is limited by the Carnot bound corresponding to the \emph{real} temperature of the WF. In the case of a phaseonium bath~\cite{scully2003extracting,scully2002extracting,rostovtsev2003extracting}, this temperature is $T_\varphi$.

\par

We analyse an Otto cycle~\cite{geva1992quantum,feldmann2004characteristics,uzdin2015quantum} for both kinds of machines (sections~\ref{sec_otto} and~\ref{sec_thermal_otto}). For machines of the first kind we then propose a modification of the Otto cycle (section~\ref{sec_modified_otto}), aimed at attaining an efficiency as high as unity, well above the Otto-cycle bound, again at the expense of mechanical work provided by the non-thermal bath. The general criteria allowing us to distinguish between the two kinds of machines are analysed (section~\ref{sec_conditions}) and the role of their quantum features is discussed (section~\ref{sec_quantum}). Our conclusions (section~\ref{sec_conclusions}) are that despite their superior performance bounds compared to standard heat engines, machines powered by non-thermal baths still adhere to the traditional rules of thermodynamics, whether or not they are powered by quantum baths or exhibit quantum features.

\section{Non-passivity in the Otto cycle: Machines of the first kind}\label{sec_otto}

\begin{figure}
  \centering
  \includegraphics[width=\columnwidth]{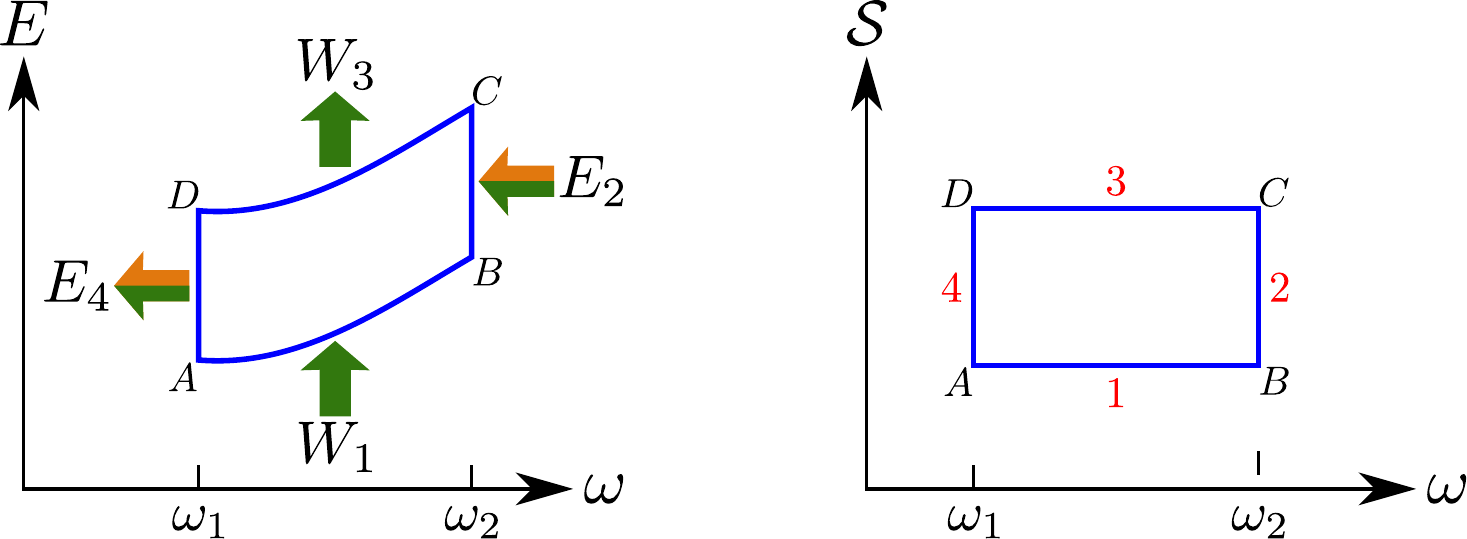}
  \caption{Representation of a quantum Otto cycle with in the frequency--energy (left) and the frequency--entropy (right) plane. Work is exchanged between the piston and the WF during the adiabatic (isentropic) strokes $1$ and $3$, in which the WF frequency is increased and decreased by the piston, respectively. In the isochoric (constant $\omega$) strokes $2$ and $4$ the WF is in contact with a ``hot'' (possibly non-thermal) or ``cold'' bath, respectively.}\label{fig_quantum_otto_cycle_1}
\end{figure}

We first revisit the analysis~\cite{abah2014efficiency} of a four-stroke quantum Otto cycle~\cite{geva1992quantum,feldmann2004characteristics,uzdin2015quantum} for thermal baths, wherein the WF is taken to be a quantised harmonic oscillator. In the isentropic strokes $1$ and $3$, the WF undergoes compression and expansion, respectively, measured by the corresponding frequency ratio $\omega_1/\omega_2\leq 1$. In the isochoric strokes $2$ and $4$, the WF is alternately coupled to an energising (``hot'') bath at temperature $T_2$ and an entropy-dump (``cold'') bath at temperature $T_1\leq T_2$, respectively. At the four end points of the Otto cycle in figure~\ref{fig_quantum_otto_cycle_1}, the respective energies of the harmonic-oscillator WF read~\cite{geva1992quantum}
\begin{subequations}\label{eq_wf_energies}
  \begin{align}
    \ew{H_1}_A&=\hbar\omega_1\left(\nbar_A+\frac{1}{2}\right)\\
    \ew{H_2}_B&=\hbar\omega_2\left(\nbar_A+\frac{1}{2}\right)\\
    \ew{H_2}_C&=\hbar\omega_2\left(\nbar_C+\frac{1}{2}\right)\\
    \ew{H_1}_D&=\hbar\omega_1\left(\nbar_C+\frac{1}{2}\right),
  \end{align}
\end{subequations}
where $H_i=\hbar\omega_i(a_i^\dagger a_i+1/2)$. Here $\nbar_{A}$ and $\nbar_C$ denote the WF's excitation number at points $A$ and $C$, respectively. For thermal baths, $\nbar_A=\nbar_1$ and $\nbar_C=\nbar_2$, where $\nbar_i$ ($i=1,2$) is the excitation number of an oscillator at temperature $T_i$ and frequency $\omega_i$. These excitation numbers are not changed in the isentropic strokes $1$ and $3$.

\par

At the heart of the analysis is the substitution of the hot thermal bath at temperature $T_2$ (coupled in the second stroke) by a non-thermal bath. In order to compare the two cycles, we restrict ourselves to a non-thermal bath generated by a unitary transformation of a thermal bath state~\cite{scully2003extracting,abah2014efficiency,rossnagel2014nanoscale,turkpence2016quantum}. This restriction allows us to relate the mean energy delivered by the non-thermal bath to its thermal counterpart. As noted above, the (originally) thermal bath causes a thermal excitation of the WF to $\nbar_2$, which we use as reference. Namely, we parameterise $\nbar_C\eqcolon \nbar_2+\Delta\nbar$, where $\Delta\nbar$ is the additional excitation of the WF when the thermal bath is replaced in the second stroke by the non-thermal one. By contrast, the cold thermal bath in stroke $4$ remains unaltered, $\nbar_A=\nbar_1$.

\par

From the WF energies~\eqref{eq_wf_energies} at the four end points of the strokes forming the cycle in figure~\ref{fig_quantum_otto_cycle_1} we can compute the energy transfer in each stroke~\cite{abah2014efficiency},
\begin{subequations}\label{eq_general_energies}
  \begin{align}
    W_1&=\ew{H_2}_B-\ew{H_1}_A=\hbar\left(\omega_2-\omega_1\right)\left(\nbar_1+\frac{1}{2}\right)\label{eq_W1}\\
    E_2&=\ew{H_2}_C-\ew{H_2}_B=\hbar\omega_2\left(\nbar_2+\Delta\nbar-\nbar_1\right)\label{eq_E2}\\
    W_3&=\ew{H_1}_D-\ew{H_2}_C=\hbar\left(\omega_1-\omega_2\right)\left(\nbar_2+\Delta\nbar+\frac{1}{2}\right)\\
    E_4&=\ew{H_1}_A-\ew{H_1}_D=\hbar\omega_1\left(\nbar_1-\nbar_2-\Delta\nbar\right).
  \end{align}
\end{subequations}
In the first and third strokes only the classical piston drives the WF so that the time evolution of the machine is unitary, hence the energy exchanged between the WF and the piston is unambiguously in the form of work and no heat transfer is involved, $Q_1=Q_3=0$. By contrast, for the strokes $2$ and $4$, wherein the WF is coupled to the (possibly non-thermal) baths, the nature of the energy exchanges $E_2$ and $E_4$ calls for further consideration, as detailed below.

\par

Under the conditions discussed in section~\ref{sec_conditions}, the non-thermal bath in stroke $2$ promotes the WF to a non-passive state, $\rho_C=U\rho_{T_2}U^\dagger$, expressed by a unitary transformation of a thermal state at temperature $T_2$ (see~\ref{app_ergotropy}). This unitary transformation necessarily increases the energy of the state, so that $\Delta\nbar>0$, because a thermal state has the lowest energy for a given entropy~\cite{schwablbook}. The WF energy at point $C$ has two distinct parts: The part proportional to $\nbar_2$ corresponds to the WF's thermal (passive) energy and the part proportional to $\Delta\nbar$ corresponds to its non-passive energy (ergotropy), which increases from zero to $\mathcal{W}_C$ on account of the bath. This ergotropy change corresponds to an ``internal work'' stored in the WF, i.e., useful energy extractable by a piston~\cite{gelbwaser2013work,gelbwaser2014heat}. This work evaluates to
\begin{equation}\label{eq_W2}
  W_2\equiv\mathcal{W}_C=\ew{H_2}_C-\ew{H_2}_C^\mathrm{therm}=\hbar\omega_2\Delta\nbar>0,
\end{equation}
which is the difference between the WF energy and its passive (thermal) counterpart~\cite{allahverdyan2004maximal,binder2015quantum}.

\par

The heat transferred in the second stroke,
\begin{equation}\label{eq_Q2}
  Q_2\coloneq E_2-W_2=\hbar\omega_2\left(\nbar_2-\nbar_1\right),
\end{equation}
is the same that a thermal bath at temperature $T_2$ would provide. Hence, such a machine is powered by both \emph{work and heat} from the non-thermal bath. Similarly, the energy $E_4$ exchanged in the fourth stroke can be divided into $W_4=-\hbar\omega_1\Delta\nbar$ and $Q_4=\hbar\omega_1(\nbar_1-\nbar_2)$. 

\par

This division of the energies $E_2$ and $E_4$ (exchanged between the WF and the bath) into work and heat (for a non-passive WF state) is one of our central results that bears important operational consequences.

\par

The efficiency of this machine, as of any engine, is the ratio of the work output to the input energy~\cite{schwablbook}, 
\begin{equation}\label{eq_eta_general} 
  \eta\coloneq\frac{-W}{E_\mathrm{in}},
\end{equation}
which holds as long as $W\leq0$, i.e., work is extracted by the piston. Substituting the energies~\eqref{eq_general_energies} yields the same expression as for a thermal bath~\cite{geva1992quantum,abah2014efficiency},
\begin{equation}\label{eq_general_eta} 
  \eta=\frac{-(W_1+W_3)}{E_2}=1-\frac{\omega_1}{\omega_2}.
\end{equation}
However, the bounds on the efficiency~\eqref{eq_general_eta} may be strongly modified by the WF's non-passivity. To derive these bounds, we require the work to be non-positive (as fits an engine), implying $\nbar_1-(\nbar_2+\Delta\nbar)\leq0$. For thermal baths ($\Delta\nbar=0$) this condition reduces to the Carnot bound,
\begin{equation}
  \frac{\omega_1}{\omega_2}\geq \frac{T_1}{T_2}\quad\Rightarrow\quad\eta\leq1-\frac{T_1}{T_2}\equiv\eta_\mathrm{Carnot}.
\end{equation}
The Carnot bound has here been obtained from the argument that an engine (by definition) has to provide work. This \emph{thermodynamic} bound also follows from applying the first and second laws of thermodynamics to a cyclic heat engine~\cite{schwablbook}.

\par

For $\Delta\nbar>0$ we may parameterise the WF by an excitation parameter that has the appearance of a \emph{fictitious} ``temperature'' $\Thetaf>T_2$, defined as
\begin{equation}\label{eq_thetaf}
  \nbar_2+\Delta\nbar\eqcolon\left[\exp\left(\frac{\hbar\omega_2}{\kB\Thetaf}\right)-1\right]^{-1}.
\end{equation}
Here $\Thetaf$ only parameterises the WF's excitation, but \emph{not} its state, which is \emph{non-passive} and hence \emph{non-thermal}. The condition on work extraction then yields
\begin{equation}\label{eq_eta}
  \frac{\omega_1}{\omega_2}\geq \frac{T_1}{\Thetaf}\quad\Rightarrow\quad\eta\leq1-\frac{T_1}{\Thetaf},
\end{equation}
which implies an apparent violation of the Carnot bound for a non-passive WF, since $1-T_1/\Thetaf>1-T_1/T_2$.

\par
\begin{figure}
  \centering
  \includegraphics[width=\columnwidth]{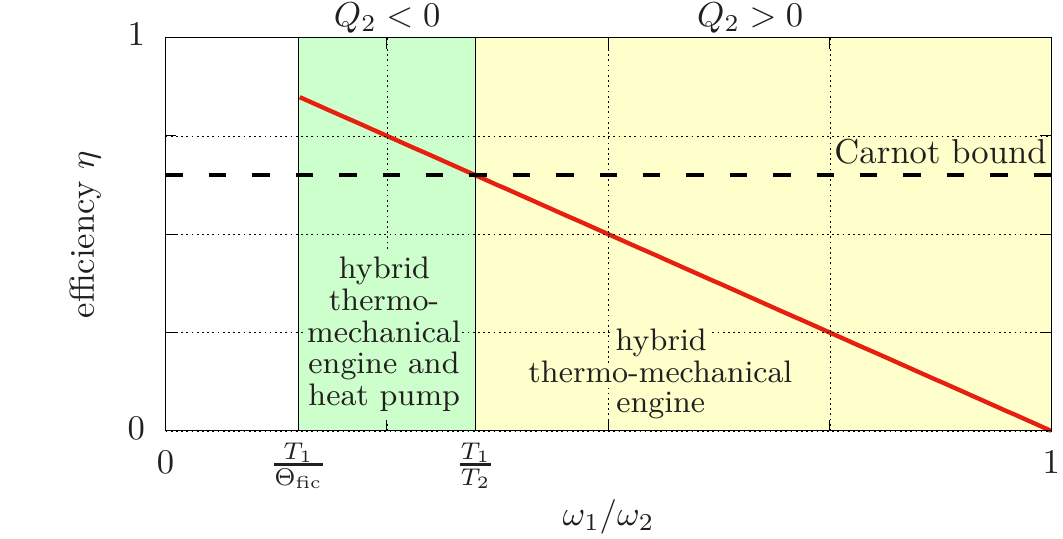}
  \caption{Efficiency~\eqref{eq_general_eta} of the non-thermal Otto machine as a function of the frequency ratio $\omega_1/\omega_2$ for a non-passive WF. In the right sub-Carnot (yellow) region (thermo-mechanical engine) the non-thermal bath provides heat and work. In the central super-Carnot (green) region the WF dumps heat into the ``hot'' bath as a heat pump, but the machine still provides work. For even smaller frequency ratios (left white area) the machine is not an engine.}\label{fig_efficiency_nonthermal_1}
\end{figure}
\par

\par
\begin{figure*}
  \centering
  \includegraphics[width=1.2\columnwidth]{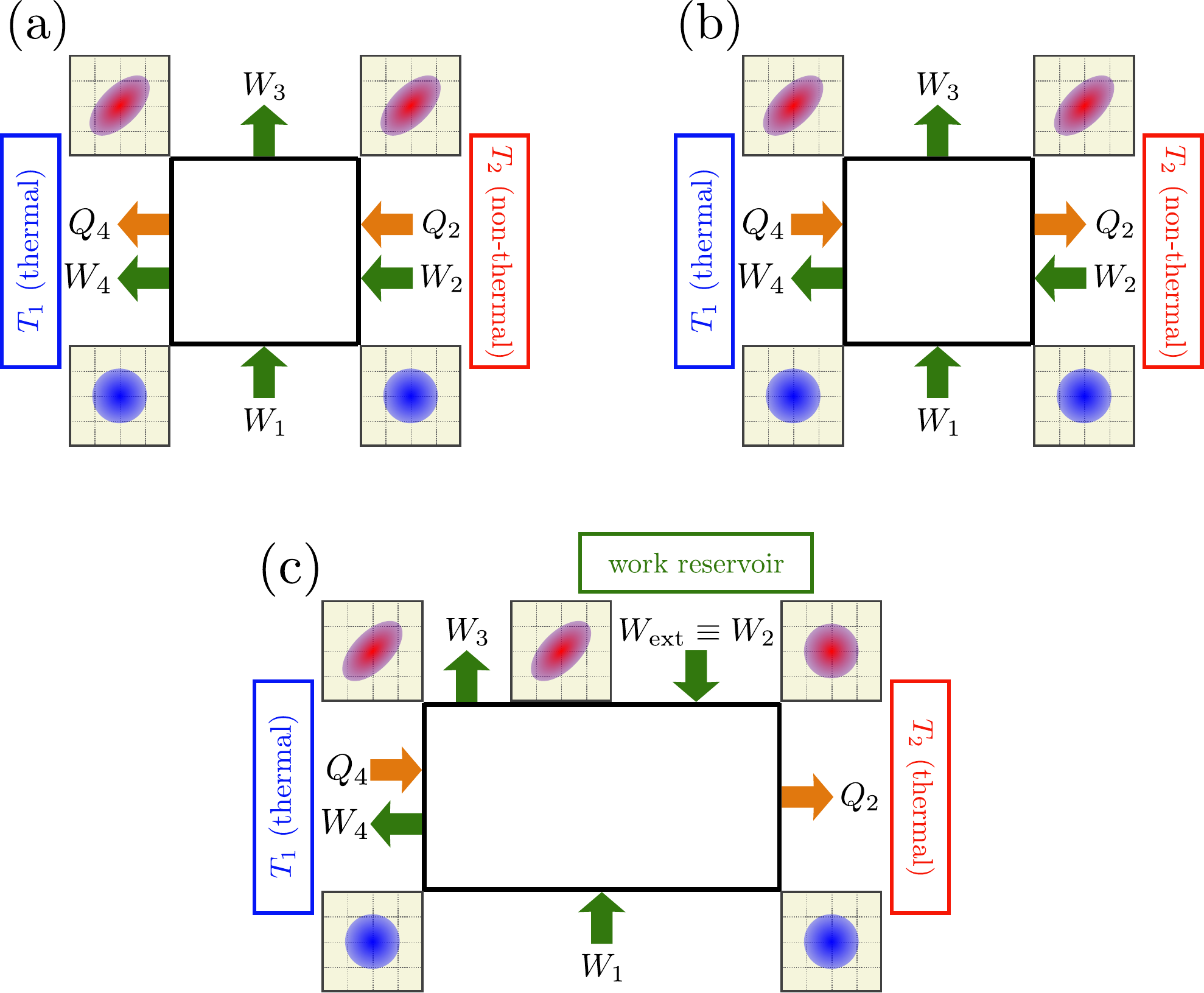}
  \caption{Otto cycle involving a squeezed thermal bath. (a) In the parameter range that yields sub-Carnot efficiency, corresponding to the yellow right region in figure~\ref{fig_efficiency_nonthermal_1}, the non-thermal bath provides work and heat to the WF. Centred circles indicate passive (thermal) states of the WF and centred ellipses squeezed thermal states. (b) The same Otto cycle in the parameter range that yields super-Carnot efficiency, corresponding to the green central region in figure~\ref{fig_efficiency_nonthermal_1}. In this range the engine also acts as a heat pump. Note the reversed arrows of the heat flows ($Q_2$ and $Q_4$): Whilst the non-thermal bath still provides work, in this range the heat is provided by the nominally ``cold'' bath. (c) Mapping of the Otto cycle onto an equivalent thermo-mechanical machine involving an external work source provided by a device (not the piston). Here we show this mapping in the super-Carnot range, but the same holds in the sub-Carnot range by simply reversing the arrows of the heat flows ($Q_2$ and $Q_4$).}\label{fig_quantum_otto_cycle_2_new}
\end{figure*}
\par

Let us examine this alleged violation (figure~\ref{fig_efficiency_nonthermal_1}). The ``hot'' non-thermal bath only provides heat $Q_2\geq0$ as long as $\nbar_2\geq\nbar_1$, corresponding to the ``sub-Carnot'' range $\omega_1/\omega_2\geq T_1/T_2$ (figure~\ref{fig_quantum_otto_cycle_2_new}a). In the ``super-Carnot'' range $T_1/\Thetaf\leq\omega_1/\omega_2\leq T_1/T_2$ this bath does not provide any heat to the WF. On the contrary, in this range $\nbar_2\leq\nbar_1$, causing the WF to dump heat into the nominally ``hot'' bath, we find that $Q_2\leq0$ (figure~\ref{fig_quantum_otto_cycle_2_new}b). The hybrid machine then acts as an engine but also pumps heat $Q_4\geq0$ out of the ``cold'' bath, along with work dumping $W_4\leq0$ into this bath. Still, the negativity of $E_4$ in the super-Carnot range can make this heat pump distinct from a refrigerator (\ref{app_refrigerator}). The input energy $E_2$ is then still positive, owing to the work contribution $W_2=E_2+|Q_2|$ [equation~\eqref{eq_W2}] endowed by the non-thermal bath's ergotropy. The first and second laws~\cite{schwablbook} hold in both the sub- and super-Carnot ranges under the conditions
\begin{subequations}\label{eq_first_second_law}
  \begin{equation}
    W_1+W_2+W_3+W_4+Q_2+Q_4=0
  \end{equation}
  and
  \begin{equation}\label{eq_first_second_law_2}
    \frac{Q_2}{T_2}+\frac{Q_4}{T_1}\leq 0.
  \end{equation}
\end{subequations}
Here we have again relied on our assumption that the non-thermal bath is unitarily generated from a thermal bath, hence equation~\eqref{eq_first_second_law_2} represents the equilibrium version of the second law (see~\ref{app_second_law}).

\par

We have arrived at a crucial result: As opposed to the genuine Carnot bound, the efficiency bound in equation~\eqref{eq_eta} \emph{cannot} be derived from the laws of equilibrium thermodynamics that are expressed by equations~\eqref{eq_first_second_law}: $\Thetaf$ is \emph{not} a temperature and hence does not appear in equation~\eqref{eq_first_second_law_2}. Namely, although the efficiency~\eqref{eq_eta_general} can only take values up to $1$ by the first law of thermodynamics (energy conservation), the \emph{maximal} efficiency~\eqref{eq_eta} for given parameters \emph{does not} follow from the second-law statement~\eqref{eq_first_second_law_2} for this equilibrium scenario: This second-law statement only restricts the heat transfer, but not the work imparted by the non-thermal bath, because this work is transferred via a unitary transformation of the WF state, which is an isentropic process. The bound~\eqref{eq_eta} is thus \emph{not} a thermodynamic bound. We shall revisit this result below.

\par

As follows from the discussion above, since this machine is fed by mechanical work, its efficiency [equation~\eqref{eq_general_eta}] is bounded by equation~\eqref{eq_eta}, rather than by the Carnot bound that only applies to heat engines (figure~\ref{fig_efficiency_nonthermal_1}). Whereas the energy conversion efficiency of a heat engine is limited by $W_\mathrm{out}\leq \eta_\mathrm{Carnot}Q_\mathrm{in}$, its mechanical-motor counterpart is bounded by the input work, $W_\mathrm{out}\leq W_\mathrm{in}$. Hence, machines powered by coherent (or squeezed) baths realise the regime of hybrid thermo-mechanical operation (where heat and work supplied by the bath are converted into work) that lies between the heat- and mechanical-engine regimes.

\par

The difference between a standard heat engine (which, by definition, is energised exclusively by heat) and this hybrid thermo-mechanical machine of the first kind becomes apparent in the extreme case $T_1=T_2=0$. A machine of the first kind can then still deliver work,
\begin{equation}\label{eq_work_zero_temperature}
  W=-\hbar(\omega_2-\omega_1)\Delta\nbar<0,
\end{equation}
although no heat is imparted to the WF by the (pure-state) bath. The machine is then an \emph{entirely mechanical} engine, energised by $E_2=\hbar\omega_2\Delta\nbar>0$, which is exclusively work transfer from the zero-temperature bath, since the WF state becomes pure and non-passive (e.g., squeezed vacuum) as a result of the WF--bath interaction. Namely, for a pure state of both the cold and the hot (non-thermal) baths, the evolution of the WF is unitary, so that the WF energy increase in the second stroke, wherein the WF is coupled to the non-thermal bath, is \emph{isentropic} and has nothing to do with heat transfer from the bath.

\par

We may replace the non-thermal Otto cycle (figures~\ref{fig_quantum_otto_cycle_2_new}a and~\ref{fig_quantum_otto_cycle_2_new}b) by an equivalent cycle (figure~\ref{fig_quantum_otto_cycle_2_new}c) involving a hot thermal reservoir at temperature $T_2$ and an external work source (which is not the piston). After the second stroke this external device performs a unitary transformation on the thermal WF state that promotes it to the same non-passive state it would have attained via contact with a non-thermal bath. The amount of work $W_\mathrm{ext}$ invested by this device is the same as $W_2$, the work stored in the WF state that a non-thermal bath would have provided. This equivalent cycle demonstrates the \emph{hybrid thermo-mechanical} nature of the engine, since the non-thermal bath is generated by a (work-investing) unitary transformation of a thermal state (see \ref{app_mapping} and reference~\cite{ekert1990canonical}). The equivalence of this cycle and the non-thermal Otto cycle follows from our analysis at the beginning of this section, where we found that the heat $Q_2$ [equation~\eqref{eq_Q2}] provided by the non-thermal bath is the \emph{same} as the heat that a thermal bath at temperature $T_2$ would have provided. The energy surplus imparted by the non-thermal bath was identified to be the work $W_2$ in equation~\eqref{eq_W2}. This equivalence supports our conclusion that the maximum efficiency~\eqref{eq_eta} is not a thermodynamic bound---the work $W_\mathrm{ext}\equiv W_2$ imparted by the auxiliary work reservoir is not bounded by the second law of thermodynamics, whereas the heat exchanges are.

\par

Our analysis can be illustrated for a \emph{squeezed thermal bath}~\cite{kim1989properties,marian1993squeezed,rossnagel2014nanoscale}, for which the oscillator WF evolves into a (non-passive) squeezed thermal state owing to work and heat imparted by the bath. The deviation of the WF's excitation number from thermal equilibrium is
\begin{equation}
  \Delta\nbar=(2\nbar_2+1)\sinh^2(r)>0,
\end{equation}
with $r>0$ denoting the squeezing parameter. At high temperature $T_2$,
\begin{equation}\label{eq_thetaf_squeezed}
  \Thetaf=T_2\cosh(2r)>T_2,
\end{equation}
but $\Thetaf$ should \emph{not} be mistaken for a temperature, as stressed above.

\par

An alternative is a \emph{coherent thermal state} of the bath~\cite{oz1991thermal}, which, in turn, yields a coherent thermal state of the WF, represented in phase space by a Gaussian displaced by $\alpha$, for which $\Delta\nbar=|\alpha|^2$. At high $T_2$ we then find
\begin{equation}
  \Thetaf=T_2+\frac{\hbar\omega_2}{\kB}|\alpha|^2.    
\end{equation}
The energy obtained from such a bath is clearly a combination of heat and work: The master equation for a WF in contact with this bath contains a thermalising Liouvillian term and a Hamiltonian term (``cavity pump'') that generates coherent displacement~\cite{carmichaelbookstatistical2,gardinerbook,ritsch2013cold}.

\section{Simultaneous engine and refrigerator action in a modified Otto cycle in machines of the first kind}\label{sec_modified_otto}

\begin{figure}
  \centering
  \includegraphics[width=0.9\columnwidth]{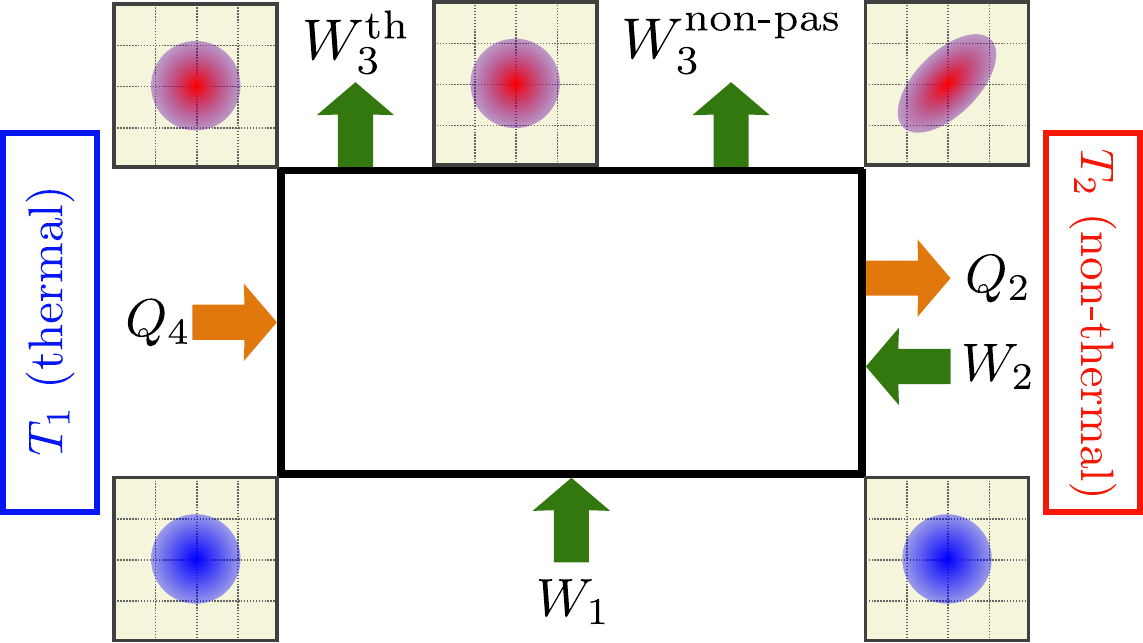}
  \caption{Modification of the Otto cycle yielding higher efficiency: After the non-thermal bath promotes the WF to a non-passive state, the piston performs a reverse unitary transformation to extract the work $W_3^\text{non-pas}$. In the subsequent isentropic stroke the WF frequency is reduced from $\omega_2$ to $\omega_1$, extracting the work $W_3^\mathrm{th}$. The figure (like figure~\ref{fig_quantum_otto_cycle_2_new}b) shows the operation for $\nbar_2<\nbar_1$, where the machine acts simultaneously as an engine and a refrigerator.}\label{fig_quantum_otto_cycle_3}
\end{figure}

Potentially extractable work is lost in the Otto cycle because the ergotropy stored by a non-passive WF is dumped into the ``cold'' bath in stroke $4$. To avoid this loss of extractable work and make the stored ergotropy useful, we suggest to modify the Otto cycle as follows: Before the adiabatic stroke $3$ the piston will perform on the WF the inverse of the unitary transformation that rendered the WF non-thermal in stroke $2$ (e.g., the inverse of the squeezing or displacement transformations). This (ideally cost-free) operation will release the excess ergotropy of the WF and transform it back to a passive state. After this unitary transformation, the WF undergoes the same adiabatic frequency change as in the standard Otto cycle. Note, however, that the order of the two actions in stroke $3$ can be arbitrary. The work extracted by the piston in this modified stroke (figure~\ref{fig_quantum_otto_cycle_3}) is
\begin{equation}
  W_3^\prime=W_3^\mathrm{th}+W_3^\text{non-pas}=\hbar\left(\omega_1-\omega_2\right)\left(\nbar_2+\frac{1}{2}\right)-\hbar\omega_2\Delta\nbar,
\end{equation}
where $W_3^\mathrm{th}$ and $W_3^\text{non-pas}=-W_2$ denote work extraction after and before the transformation, respectively. The last term on the r.h.s.\ is the ergotropy obtained from the non-thermal bath, but with a negative sign.

\par

\begin{figure}
  \centering
  \includegraphics[width=\columnwidth]{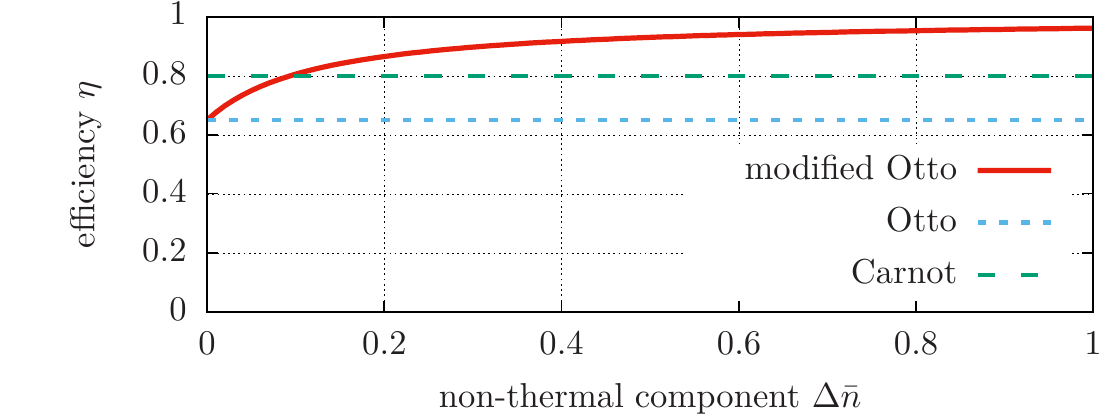}
  \caption{Efficiency~\eqref{eq_efficiency_modified_otto} of the modified Otto cycle that exploits the non-passivity of the WF as a function of the non-thermal component $\Delta\nbar$ in the regime $\nbar_2\geq\nbar_1$. Parameters (in an arbitrary energy unit): $\hbar\omega_1=7$, $\hbar\omega_2=20$, $\kB T_1=2$ and $\kB T_2=10$. In the range $\nbar_2<\nbar_1$ the efficiency of the machine is unity.}\label{fig_efficiency_nonthermal_2}
\end{figure}

In the parameter range $\nbar_2\geq\nbar_1$ the modified Otto cycle represents a thermo-mechanical engine that dumps heat $Q_4\leq0$ into the ``cold'' bath, and equation~\eqref{eq_eta_general} yields the efficiency (figure~\ref{fig_efficiency_nonthermal_2})
\begin{equation}\label{eq_efficiency_modified_otto}
  \eta=\frac{-(W_1+W_3^\prime)}{E_2}=1-\frac{(\nbar_2-\nbar_1)\omega_1}{(\nbar_2+\Delta\nbar-\nbar_1)\omega_2}\geq 1-\frac{\omega_1}{\omega_2}.
\end{equation}

\par
\par

If, however, $\nbar_2<\nbar_1$, the machine acts simultaneously as a thermo-mechanical engine and a refrigerator, i.e., it extracts heat $Q_4>0$ from the nominally ``cold'' bath (\ref{app_refrigerator}). Hence, the input energy in equation~\eqref{eq_eta_general} is then $E_\mathrm{in}=E_2+Q_4$, yielding the maximal efficiency $\eta=1$. Thus, in this regime the machine not only operates as the most efficient engine possible, but, surprisingly, also refrigerates the ``cold'' bath. The heat extraction $Q_4$ from the cold bath is the same as for the \emph{thermal} Otto refrigerator, i.e., the refrigerator obtained by inverting the Otto cycle in figure~\ref{fig_quantum_otto_cycle_1} for thermal baths. Note, however, that our dual action machine operates in the regime $\nbar_2<\nbar_1$, whereas a thermal refrigerator requires $\nbar_2>\nbar_1$. The coefficient of performance (COP) of this refrigerator, following the standard definition~\cite{schwablbook}, reads
\begin{equation}\label{eq_cop}
  \mathrm{COP}\coloneq\frac{Q_4}{W_\mathrm{inv}}=\frac{\omega_1}{\omega_2-\omega_1}\leq\frac{T_1}{T_2-T_1}\equiv\mathrm{COP}_\mathrm{Carnot},
\end{equation}
where $W_\mathrm{inv}=W_1+W_2+W_3^\prime$ denotes the net invested work (see figure~\ref{fig_quantum_otto_cycle_3}). This COP has the standard form despite an unusual feature: $W_2$ is counted here in the net invested work $W_\mathrm{inv}$, even though $W_2$ is imparted by the bath ``for free''. However, if the non-thermal bath is decomposed into a thermal bath and a work reservoir (as in figure~\ref{fig_quantum_otto_cycle_2_new}), the COP~\eqref{eq_cop} regains its traditional meaning: The unitary transformation between the strokes $2$ and $3$ produces the work $-W_2$ due to the ergotropy obtained by the WF in the second stroke. The machine then invests the work $W_\mathrm{inv}$ for refrigeration. Since the net work produced by the engine is $W_1+W_3^\prime$ [\cf equation~\eqref{eq_efficiency_modified_otto}], we obtain the work balance equation $-W_2+W_\mathrm{inv}=W_1+W_3^\prime$, which yields the denominator in equation~\eqref{eq_cop}.

\par

Thus, full exploitation of the WF's non-passivity (as in a quantum battery~\cite{alicki2013entanglement,binder2015quantacell}) can increase the machine efficiency, and allow for simultaneous production of work and refrigeration.

\section{A thermal working fluid: Machines of the second kind}\label{sec_thermal_otto}

\begin{figure*}
  \centering
  \includegraphics[width=1.2\columnwidth]{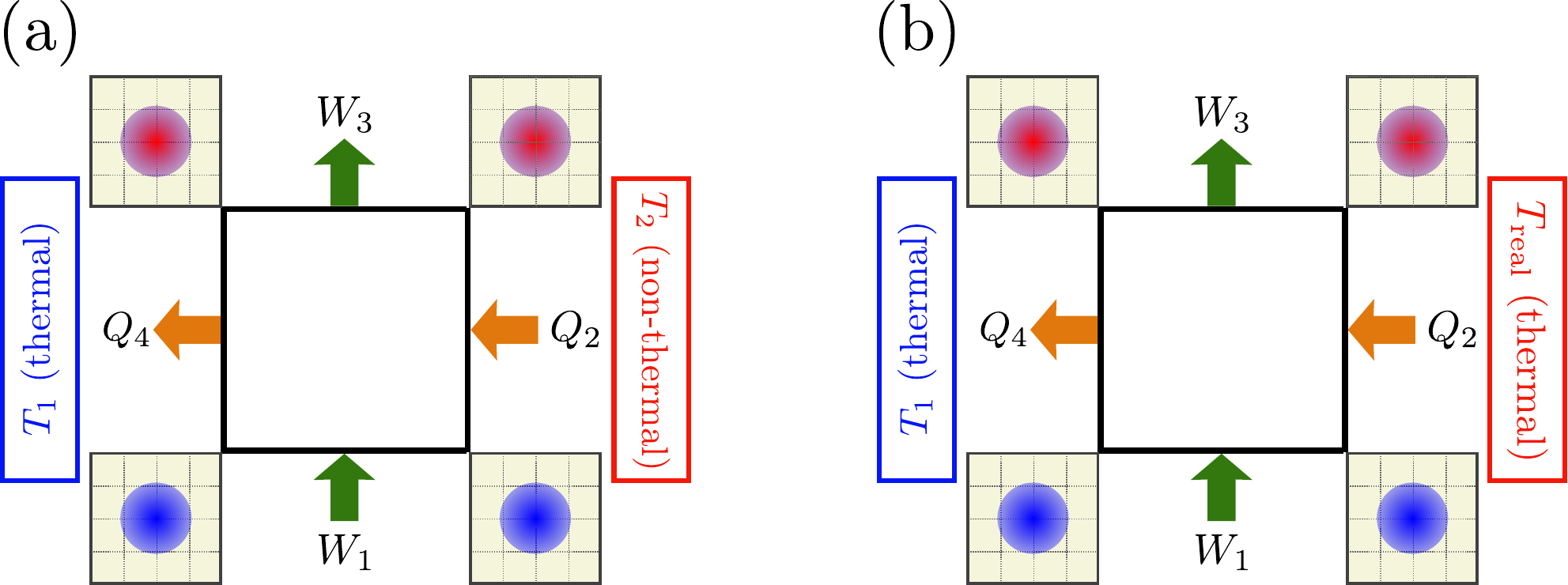}
  \caption{A heat machine with thermal WF powered by (a) a non-thermal bath with temperature parameter $T_2$ is equivalent to a heat machine powered by (b) a thermal bath at (real) temperature $\Treal>T_2$.}\label{fig_quantum_otto_cycle_4}
\end{figure*}

We have thus far considered machines of the first kind in which the WF draws both work and heat from the non-thermal reservoir, promoting it to a non-passive state. However, there are machines wherein the WF is thermalised by the non-thermal bath (see section~\ref{sec_conditions}), so that no work is imparted by the bath, despite it being non-thermal. This means that the excitation $\nbar_2+\Delta\nbar$ in equations~\eqref{eq_general_energies} corresponds to a \emph{real} temperature $\Treal$ of the WF~\cite{alicki2015nonequilibrium} (instead of the parameter $\Thetaf$ used for mere convenience). The first and second laws then read [in contrast to equations~\eqref{eq_first_second_law}]
\begin{subequations}\label{eq_first_second_law_thermal_wf}
\begin{equation}\label{eq_first_second_law_thermal_wf_1}
  W_1+W_3+\mathcal{Q}_2+\mathcal{Q}_4=0
\end{equation}
and
\begin{equation}\label{eq_first_second_law_thermal_wf_2}
  \frac{\mathcal{Q}_2}{\Treal}+\frac{\mathcal{Q}_4}{T_1}\leq 0,
\end{equation}
\end{subequations}
where $\mathcal{Q}_2\coloneq E_2$ and $\mathcal{Q}_4\coloneq E_4$ with the energies $E_i$ being defined in equations~\eqref{eq_general_energies}. The notation $\mathcal{Q}_i$ for the heat in stroke $i$ is meant to distinguish them from $Q_i$ considered in section~\ref{sec_otto}.

\par

Consequently, in this regime the machine operates as a genuine heat engine (see figure~\ref{fig_quantum_otto_cycle_4}) whose efficiency is restricted by the Carnot bound
\begin{equation}\label{eq_carnot_thermal_wf}
  \eta=\frac{-(W_1+W_3)}{\mathcal{Q}_2}\leq1-\frac{T_1}{\Treal}\equiv\eta_\mathrm{Carnot}
\end{equation}
corresponding to this real temperature. The (``original'') temperature $T_2$ of the bath, prior to the unitary transformation, plays no role; the only temperatures of consequence are $T_1$ and $\Treal$. These are the temperatures that appear in the second-law expression~\eqref{eq_first_second_law_thermal_wf_2}, which, together with the first law~\eqref{eq_first_second_law_thermal_wf_1}, gives rise to the Carnot bound~\eqref{eq_carnot_thermal_wf}. For a harmonic oscillator WF, this regime is realised, e.g., for a cavity being fuelled by a phaseonium bath where $\Treal=T_\varphi$~\cite{scully2003extracting,scully2002extracting,rostovtsev2003extracting}, its $N$-level generalisation~\cite{turkpence2016quantum}, or by a beam of thermal entangled atoms~\cite{dillenschneider2009energetics}. Note that contrary to machines of the first kind treated in section~\ref{sec_otto}, in machines of the second kind $\Delta\nbar$ may, in principle, become negative, thereby decreasing $\nbar_C$ compared to $\nbar_2$ in equations~\eqref{eq_wf_energies} and~\eqref{eq_general_energies}. This case is exemplified by a phaseonium bath with the wrong choice of phase~$\varphi$~\cite{scully2003extracting}. Only if $\Delta\nbar>0$ is $\Treal>T_2$.

\par

We have seen in section~\ref{sec_otto} that in the regime of a non-passive WF state the engine is rendered purely mechanical and capable of providing work for $T_1=T_2=0$ [equation~\eqref{eq_work_zero_temperature}]. Is this behaviour changed when the WF remains thermal? Close inspection of the baths~\cite{scully2003extracting,scully2002extracting,rostovtsev2003extracting,dillenschneider2009energetics,turkpence2016quantum} (which are all generated by atomic beams) that thermalise the WF to $\Treal$ reveals that $T_2=0$ entails $\Treal=0$, hence no work is generated by a thermal WF in the limit of pure bath states. However, this statement must be examined for each specific bath. For instance, for a cavity-mode WF interacting with an atomic-beam bath, the cavity mode may attain a finite temperature $\Treal>0$ even if the atoms are originally at zero temperature, i.e., in a pure state~\cite{dag2016multiatom}. This is due to the fact that the atoms in the bath are removed and traced out after each interaction, thereby increasing the WF entropy.

\par

To sum up, in machines of the second kind the WF remains thermal, namely its bath-induced evolution is governed by a Gibbs-preserving map. We then recover the traditional heat-engine operation~\cite{schwablbook,alicki1979quantum}. The hybrid regime realised in machines of the first kind (sections~\ref{sec_otto} and~\ref{sec_modified_otto}) does not arise in machines of the second kind, as no work (ergotropy) is exchanged with the bath when the map is Gibbs-preserving.

\section{Conditions for working fluid thermalisation or non-thermalisation}\label{sec_conditions}

Whether or not the WF thermalises depends on the possible change of the WF--bath interaction Hamiltonian in the interaction picture, $H_\mathrm{int}(t)$, under the unitary operation $\UB$ that transforms the bath from a thermal to a non-thermal state. As detailed in~\ref{app_conditions}, if the master equation~\cite{gardinerbook} derived for the original $H_\mathrm{int}(t)$ and a thermal bath yields a thermal WF state, then also the master equation derived following the transformation $\UB$ will yield a thermal state, provided the transformed Hamiltonian
\begin{equation}\label{eq_app_conditions_UHU}
  H_\mathrm{int}(t)\mapsto (\mathds{1}_\mathrm{WF}\otimes\UB^\dagger) H_\mathrm{int}(t)(\mathds{1}_\mathrm{WF}\otimes\UB),
\end{equation}
where $\mathds{1}_\mathrm{WF}$ is the unity operator acting on the WF, \emph{retains its original form} (apart from possible renormalisations of the WF--bath coupling strengths, as, e.g., in~\cite{scully2003extracting,zubairy2002photo}). The WF will then thermalise to some real temperature, which may differ from the original bath temperature.

\par

If, however, the transformed Hamiltonian \emph{changes its form}, as, e.g., in the engines discussed in~\cite{rossnagel2014nanoscale,hardal2015superradiant}, the WF may be driven into a non-passive steady state. Hence, a change in the form of $H_\mathrm{int}(t)$ under transformation~\eqref{eq_app_conditions_UHU} is a necessary (but not sufficient) condition for a non-passive WF steady state.

\par

Physically, any interaction of the WF with the bath that causes energy exchange leads to thermalisation, provided that the WF is not initially in a dark state~\cite{scullybook}. By contrast, parametric processes, in which the energy supplied by an undepleted pump compensates for the WF--bath energy exchange (so that no energy flows between them), are the key to the formation of a non-passive steady state of the WF.

\par

One such process involves bath squeezing, described by operator $S_\mathrm{B}$, for which
\begin{equation}\label{eq_Hint_general}
  H_\mathrm{int}(t)=\sum_kg_kab_k^\dagger e^{-i\omega t}e^{i\omega_k t}+\mathrm{H.c.}
\end{equation}
is transformed into
\begin{multline}
  (\mathds{1}_\mathrm{WF}\otimes S_\mathrm{B}^\dagger) H_\mathrm{int}(t)(\mathds{1}_\mathrm{WF}\otimes S_\mathrm{B})\\=\sum_kg_ka(u_k b_k+v_k b_k^\dagger)e^{-i\omega t}e^{i\omega_k t}+\mathrm{H.c.}
\end{multline}
Here $k$ labels the bath modes, $g_k$ are the coupling constants to these modes and $u_k$ and $v_k$ describe the effect of squeezing~\cite{breuerbook,gardinerbook}. This Hamiltonian, originally written in the rotating-wave approximation, possesses, following its transformation, terms of the form $ab_k$ and $a^\dagger b_k^\dagger$ that are now resonant thanks to the (undepleted) pump action. These terms, as opposed to $ab_k^\dagger$ and $a^\dagger b_k$, do not contribute to thermalisation since they do not involve net energy exchange of the WF and the bath. This feature allows for non-passive steady states of the WF. We note that although squeezed baths in general give rise to a non-passive WF state, there exists one known exception, namely the two-level WF~\cite{breuerbook,gardinerbook,huang2012effects}.

\par

Another parametric process involves a coherent-pumping transformation, described by the displacement operator $D_\mathrm{B}$~\cite{wallsbook}. The Hamiltonian~\eqref{eq_Hint_general} is then transformed into
\begin{multline}
  (\mathds{1}_\mathrm{WF}\otimes D_\mathrm{B}^\dagger) H_\mathrm{int}(t)(\mathds{1}_\mathrm{WF}\otimes D_\mathrm{B})\\=\sum_kg_ka(b_k^\dagger e^{-i\omega t}e^{i\omega_k t}+\delta_{k,K}\beta^*)+\mathrm{H.c.},
\end{multline}
where $\beta$ is the displacement amplitude of the mode $b_K$ that acts as a resonant pump term on the WF, rendering its steady state non-passive.

\par

A necessary, but not sufficient, condition for a state $\rho$ of the WF to be passive with respect to its free Hamiltonian $H$ is that $\rho$ and $H$ commute~\cite{lenard1978thermodynamical}, which is, e.g., not the case for squeezed thermal states: These non-equilibrium states are non-passive, although they are centred, i.e., fulfill $\ew{a\pm a^\dagger}=0$. The ability of squeezed states to provide work has also been noted in~\cite{carmichaelbookstatistical2}.

\section{Is quantumness relevant to the engine performance?}\label{sec_quantum}

We have shown that the operation mode of machines powered by non-thermal baths crucially depends on the quantum state of the WF, namely, on whether it is thermal (and thus passive) or non-passive (and thus non-thermal). However, the concept of non-passivity also exists in a purely classical context~\cite{gorecki1980passive,daniels1981passivity,daprovidencia1987variational} which motivates the question: To what extent is the performance of these machines truly affected by quantum features of the bath and (or) the WF?This question calls for the elucidation of two points: (i) What are the relevant criteria for the bath or WF quantumness? (ii) Is there a compelling link between such quantumness and the machine performance?

\par

For the machines discussed here (sections~\ref{sec_otto}--\ref{sec_thermal_otto}) the first point hinges on the quantumness of a harmonic-oscillator WF, assuming that the bath is quantum (e.g., it exhibits intermode entanglement). There is a well-established criterion whereby the harmonic-oscillator state is non-classical if its Glauber--Sudarshan $P$-function is either negative or does not exist at all (in the sense of tempered distributions), namely, it is even more singular than the Dirac $\delta$-distribution~\cite{glauber1963coherent,dodonov2002nonclassical,agarwalbookquantumoptics,carmichaelbookstatistical2,schleichbook,wallsbook}. Such non-classical states cannot be described by (semi-)classical stochastic processes.

\par

In all machines powered by non-thermal baths proposed thus far~\cite{scully2003extracting,scully2002extracting,dillenschneider2009energetics,rossnagel2014nanoscale,hardal2015superradiant,turkpence2016quantum}, the dynamics of the harmonic-oscillator WF is described by linear or quadratic operators. Consequently, since the initial state of the cycle (point $A$ in figure~\ref{fig_quantum_otto_cycle_1}) is a thermal state and therefore Gaussian~\cite{adesso2007entanglement}, its evolution, whether Hamiltonian or Liouvillian, conserves the Gaussian character of the WF state~\cite{wallquist2010single}. Hence, the only types of WF states that can emerge under such interactions are (i)~thermal states, (ii)~coherent states, (iii)~squeezed states and combinations thereof, e.g., squeezed thermal states.

\par

According to the above-mentioned criterion, a displaced and (or) squeezed thermal state with thermal photon number $\nbar$ and squeezing parameter $r>0$ is non-classical if its fluctuations are smaller than the minimum uncertainty limit), which amounts to~\cite{kim1989properties,agarwalbookquantumoptics}
\begin{equation}\label{eq_squeezed_thermal_non-classical}
  \nbar<\frac{e^{2r}-1}{2}.
\end{equation}
Realistic squeezing parameters of $r\sim0.4$~\cite{wenger2004nongaussian,su2007experimental} require $\nbar\lesssim 0.6$, i.e., a very low temperature of the ``hot'' bath. In a recent experiment, squeezing of $12.7\,\mathrm{dB}$ has been achieved~\cite{eberle2010quantum}, which amounts to $r\approx 1.46$ and $\nbar\lesssim 9$. All remaining Gaussian states are deemed to be ``classical'', meaning that the evolution of the WF may be mapped onto a (semi-)classical stochastic process~\cite{carmichaelbookstatistical2,wallsbook}, even though the bath may possess quantum properties.

\par

The possible non-classicality of the WF (according to the above definition) is, however, not reflected in the machine's operational principles, i.e., the analysis of the \emph{standard cycle} in section~\ref{sec_otto} does not discriminate between classical and non-classical states---\emph{only the WF's energy matters}. By contrast, in the \emph{modified cycle} of section~\ref{sec_modified_otto} the \emph{WF state is crucial} for the machine's operation: The additional unitary transformation must be chosen according to this state and the extracted work in the modified cycle depends explicitly on the WF's ergotropy via $\Delta\nbar$ instead of the WF's excitation via $\nbar_2+\Delta\nbar$ as in the standard cycle. Consequently, for work optimisation a fixed WF energy is best divided into a small thermal component $\nbar$ and a large mechanical component $\Delta\nbar$ that stems from ergotropy transfer and is parameterised by either $r$ or $|\alpha|^2$. For a squeezed thermal bath, such a division favours a non-classical WF [equation~\eqref{eq_squeezed_thermal_non-classical}], whereas for the displaced bath the WF remains classical for any choice of $\nbar$ and $|\alpha|^2$. This most favourable (optimal) regime of the modified cycle corresponds to an almost purely mechanical operation of the machine, enabled by the ergotropy imparted by the bath to the WF. The passive (heat) contribution should best be chosen as small as possible.

\par

Hence, only the WF state's non-passivity plays a role in the modified cycle, as it determines the hybridisation of the machine's operation mode. It is in general impossible to relate non-classicality and non-passivity: Coherent thermal states are classical but non-passive, whereas squeezed thermal states are non-passive but may be either classical or non-classical. These conclusions also hold if we lift the restriction on Gaussian states: For a given WF energy, the best performance of the modified machine is realised for a WF with the highest possible ergotropy allowed for this energy. This, however, clearly shows that such machines have little in common with ``heat'' engines. They defy the need for a thermodynamic cycle: At $T_1=T_2=0$ the modified cycle simply realises a quantum battery that is charged by the bath and discharged by the piston. 

\par

The preceding discussion has dealt with the rapport between quantumness and non-passivity of the WF. However, the bath may be truly quantum-mechanical. Yet, its quantumness only affects the parameters of the WF evolution. To decide whether or not the quantumness of the bath is useful, one should ask: Given a certain energy to modify a thermal bath, what would be its optimal unitary transformation for obtaining either the largest possible WF energy (for the standard cycle), or the largest possible ergotropy of the WF (for the modified cycle) after the second stroke? Only if this optimal transformation has either no classical counterpart (for electromagnetic-field baths) or renders the bath state coherent (for atomic-beam baths), may quantumness be considered beneficial. This question must be individually answered for each type of bath.

\par

To sum up, although the bath may be designed to exhibit quantum features, such as intermode entanglement, the \emph{operational principles} of the thermo-mechanical machine, whether being operated in the standard cycle (section~\ref{sec_otto}) or the modified cycle (section~\ref{sec_modified_otto}), do not rely on the non-classicality of the harmonic-oscillator WF state. Extracted work and efficiency of the modified cycle, on the other hand, are optimised by maximising the non-passivity (ergotropy) and minimising the passive (thermal) energy of the harmonic-oscillator WF, but are not directly determined by its non-classicality.

\section{Conclusions}\label{sec_conclusions}

To conclude, we have related the unconventional performance bounds of Otto machines powered by non-thermal baths to the quantum state of their harmonic-oscillator working fluid (WF). In order to allow for a comparison with traditional Otto engines (which are energised by a thermal bath), we have assumed that the non-thermal bath is unitarily generated from a thermal one. We have dubbed as ``machines of the first kind'' those where the WF state becomes non-passive as a result of the WF--bath interaction, and the machines act concurrently as engines and heat pumps or refrigerators. If we fully exploit the ergotropy they have received from the non-thermal bath, these machines may attain maximal efficiency $\eta=1$. This maximal efficiency cannot be solely determined by thermodynamic arguments, as the second law only limits the heat, but not the work exchanged with the non-thermal baths that has been hitherto unaccounted for.

\par

By contrast, in what we dubbed as ``machines of the second kind'' the WF is thermalised by the bath. The Carnot bound applies to such machines, which are true heat engines (as opposed to the first kind). Their WF temperature and hence their efficiency may depend on the coherence or entanglement of the bath.

\par

A key insight is that the state of the bath is unimportant if all that one can measure are the extracted work and the efficiency: A thermal bath at a \emph{real} temperature $\Treal>T_2$ would yield the same extracted work as a non-thermal one at the fictitious ``temperature'' $\Thetaf=\Treal$. However, the operation mode of the respective machines is completely different. This may be revealed by observing concurrent refrigeration and work production or by diagnosing the non-passivity of the WF state, e.g., by homodyning or tomography.

\par

Under prevalent (quadratic) WF--bath interactions the steady state of the harmonic-oscillator WF is non-classical only if it interacts with a low-temperature squeezed bath. Otherwise, the WF dynamics is described by a classical stochastic process, whose parameters, however, are determined by possible quantum features of the bath (e.g., intra-atomic coherence or entanglement between atoms). Yet, non-classicality does not directly affect the machine's performance, \emph{only non-passivity matters}. Different choices of WFs may entail different criteria for their quantumness but such choices will not change this conclusion. For example, a multipartite WF state can be non-passive, regardless of whether it is separable or entangled and hence the performance criteria in sections~\ref{sec_otto} and~\ref{sec_modified_otto} will be unaffected. This issue is an example of the subtle rapport of thermodynamic and quantum features~\cite{quan2012validity,kosloff2013quantum,friedenberger2015quantum,vinjanampathy2015quantum} which have prompted proposed reformulations of thermodynamics~\cite{lieb2000fresh,horodecki2013fundamental,skrzypczyk2014work,lostaglio2015description}.

\par

These effects are realisable in optical or optomechanical setups~\cite{purdy2013strong,hammerer2014nonclassical,asjad2014robust}. Bath squeezing can be implemented in multimode cavities by highly-nonadiabatic periodic modulations of the cavity~\cite{averbukh1994enhanced}, by rapid modulation of atomic resonance frequencies~\cite{shahmoon2013engineering}, as well as by laser interactions with trapped ions~\cite{rossnagel2014nanoscale}. A coherently displaced bath is obtainable in a laser-driven cavity~\cite{gardinerbook,ritsch2013cold}. These cavity states can also be generated by beams of entangled atoms~\cite{hardal2015superradiant,dag2016multiatom}. An interesting application of the present analysis may concern a spin bath that interacts with a harmonic-oscillator WF (e.g., a cavity mode)~\cite{allahverdyan2004bath} since collective (angular-momentum) states of a spin bath are also amenable to squeezing~\cite{agarwalbookquantumoptics}.

\par

We note that the present considerations do not account for WF--bath correlation and information-thermodynamic effects~\cite{toyabe2010experimental,deffner2013information,gelbwaser2013workextraction,goold2014nonequilibrium,horowitz2014quantum,brandner2015coherence,elouard2015reversible,gardas2015thermodynamic,goold2016role,lutz2015information,parrondo2015thermodynamics}. Nor are we concerned with bath-preparation costs~\cite{zubairy2002photo,hardal2015superradiant,turkpence2016quantum}. We are only concerned with the question: How to best exploit non-thermal baths as resources for machine operation?

\begin{acknowledgments}
We thank Robert Alicki for helpful discussions and the ISF and BSF for support. D.~G.-K. acknowledges the support of the Center for Excitonics, an Energy Frontier Research Center funded by the U.S. Department of Energy under award DE-SC0001088 (energy conversion process), and the CONACYT (non-equilibrium thermodynamics).
\end{acknowledgments}

\appendix

\section{Non-passivity, ergotropy and work storage}\label{app_ergotropy}

Ergotropy is a function of a quantum state $\rho$ and a Hamiltonian $H$ that quantifies the maximal work extractable from this state~\cite{pusz1978passive,lenard1978thermodynamical,allahverdyan2004maximal,alicki2013entanglement,binder2015quantum},
\begin{equation}
  \mathcal{W}(\rho,H)\coloneq\Tr(\rho H)-\min_U\Tr(U\rho U^\dagger H),
\end{equation}
where the minimisation is over the set of all possible unitary transformations of the Hilbert space. A state $\rho_\mathrm{pas}$ is called passive if no work can be extracted from it, i.e.,
\begin{equation}
  \Tr(\rho_\mathrm{pas} H)\leq\Tr(U\rho_\mathrm{pas} U^\dagger H)
\end{equation}
for all $U$ and therefore $\mathcal{W}(\rho_\mathrm{pas},H)=0$. Ergotropy is thus the difference of the state's energy and the corresponding passive energy,
\begin{equation}
  \mathcal{W}(\rho,H)=\Tr(\rho H)-\Tr(\rho_\mathrm{pas} H).
\end{equation}
Hence, the internal energy of a non-passive quantum state $\rho$ can be divided into passive energy and ergotropy. When two systems are in contact, work transfer between them is associated with ergotropy change. Note that while ergotropy is non-negative, work extracted from a system by a piston is negative.

\par

In equations~\eqref{eq_general_energies} (section~\ref{sec_otto}) we have used the feature that if the state $\rho$ is obtained by a (entropy-conserving) unitary transformation of a thermal state, the corresponding passive state remains this thermal state. This follows from the fact that a thermal state, which is necessarily passive~\cite{lenard1978thermodynamical}, is the minimum-energy state for a given entropy~\cite{schwablbook}.

\par

The generalisation of equations~\eqref{eq_general_energies} to an \emph{arbitrary} non-thermal bath would amount to the replacement of $\nbar_2$ by the excitation number $\nbar_\mathrm{pas}$ corresponding to the passive energy of the WF at point $C$ of the cycle (figure~\ref{fig_quantum_otto_cycle_1}). Yet, $\Delta\nbar$ would still correspond to the non-passive energy (ergotropy) obtained from the mechanical work imparted by the bath via a unitary transformation. The parameterisation of the WF energy by $\Thetaf$ [equation~\eqref{eq_thetaf}] still holds. By contrast, there is no notion of $T_2$ when the bath coupled to the WF (in the second stroke) is not generated by a unitary operation from a thermal bath. In such cases, the strokes $2$ and $4$ can no longer be regarded as coupling to ``hot'' and ``cold'' baths, but rather as an energising stroke and a state-resetting stroke (that closes the cycle).

\section{Heat-pumping and work-dumping by a non-passive working fluid}\label{app_refrigerator}

In the quantum Otto cycle powered by a non-thermal bath that renders the WF non-passive, stroke $4$ in the super-Carnot range (figure~\ref{fig_quantum_otto_cycle_2_new}b) corresponds to work dumping into the ``cold'' bath, $W_4<0$, along with heat pumping from this bath, $Q_4>0$. The effect of such work dumping is subtle: Typically, this work corresponds to a coherent excitation (displacement) of a bath mode~\cite{gelbwaser2014heat} that may be extracted by an appropriate piston. If this work is not extracted, it may (in the case of a finite heat capacity) heat up this bath and thereby counter the heat pumping out of it.

\par

By contrast, the modified Otto cycle (figure~\ref{fig_quantum_otto_cycle_3}), where the work dumping in stroke $4$ is absent, gives rise to genuine ``cold''-bath refrigeration in the super-Carnot range.

\section{Second law for non-thermal baths and a non-passive working-fluid}\label{app_second_law}

Consider what we refer to as machines of the first kind, wherein the second stroke (which describes the interaction with a non-thermal bath) transforms the WF into some non-passive state. According to~\ref{app_ergotropy}, the WF state at point $C$ (figure~\ref{fig_quantum_otto_cycle_1}) can be written (for an arbitrary non-thermal bath) as $\rho_C=U\rho_\mathrm{pas}U^\dagger$, i.e., as a unitarily-transformed passive state. Consequently, the second stroke can be thought of as consisting of a stroke that drives the WF into the passive state $\rho_\mathrm{pas}$ (via heat exchange and an increase of the WF's entropy, $\Delta S_2>0$) and a subsequent isentropic ergotropy-transfer stroke that is associated to work performed by the bath via the unitary transformation $U$, leading to the state $\rho_C$.

\par

In our study, the passive state is an \emph{equilibrium} Gibbs state. Hence, the entropy exchange in this stroke is given by the standard (equilibrium) form of the second law, $\Delta S_2\geq Q_2/T_2$.

\section{Equivalence of non-thermal baths and external work sources}\label{app_mapping}

The (Lindblad) master equation (ME) in the interaction picture for a bosonic WF in contact with a phase-sensitive unitarily-transformed bath (from a thermal to a non-thermal state) with Liouvillian $\mathcal{L}_\text{non-th}$ and initial condition $\rho_0$,
\begin{equation}\label{eq_master_nonthermal}
  \dot\rho(t)=\mathcal{L}_\text{non-th}\rho(t)\mathrm{\quad with \quad} \rho(0)=\rho_0,
\end{equation}
is unitarily equivalent~\cite{ekert1990canonical} to a ME with Liouvillian $\mathcal{L}_\mathrm{th}$, i.e., assuming a thermal state of the bath,
\begin{equation}\label{eq_master_thermal}
  \dot\rho_\mathrm{th}(t)=\mathcal{L}_\mathrm{th}\rho_\mathrm{th}(t)\quad\mathrm{with}\quad \rho_\mathrm{th}(0)=\Uwf^\dagger\rho_0\Uwf.
\end{equation}
Here the WF state $\rho_\mathrm{th}(t)$ signifies the solution of the ME with a thermal bath. When in contact with a non-thermal bath, the WF state can then be obtained by unitarily transforming (e.g., squeezing) the WF state determined by equation~\eqref{eq_master_thermal}, 
\begin{equation}\label{eq_master_solution_u}
  \rho(t)=\Uwf\rho_\mathrm{th}(t)\Uwf^\dagger.
\end{equation}
For the Otto cycle depicted in figure~\ref{fig_quantum_otto_cycle_1} only the steady-state solution of equation~\eqref{eq_master_nonthermal} is required. Hence, figures~\ref{fig_quantum_otto_cycle_2_new}a and~\ref{fig_quantum_otto_cycle_2_new}b correspond to obtaining the steady-state solution of equation~\eqref{eq_master_nonthermal}, while figure~\ref{fig_quantum_otto_cycle_2_new}c describes the equivalent procedure of finding the thermal steady-state solution of the ME~\eqref{eq_master_thermal} (which is independent of the initial condition) and unitarily transforming it subsequently according to equation~\eqref{eq_master_solution_u}.

\section{Transformations of the interaction Hamiltonian that generate non-thermal baths}\label{app_conditions}

Upon tracing out the bath, $\rho_\mathrm{WF}\coloneq\Tr_\mathrm{B}(\rho)$, the solution of the von Neumann equation in the interaction picture involving a thermal bath yields the WF state
\begin{equation}\label{eq_app_conditions_me1}
  \rho_\mathrm{WF}(t)=\Tr_\mathrm{B}\left[U(t,0;H_\mathrm{int})\left(\rho_\mathrm{WF}(0)\otimes\rho_\mathrm{B}^\mathrm{th}\right)U^\dagger(t,0;H_\mathrm{int})\right].
\end{equation}
Here $U(t,0;H_\mathrm{int})\coloneq T\exp(-i\int_0^tH_\mathrm{int}(\tau)\dd\tau/\hbar)$, $T$ being the time-ordering operator, denotes the time evolution operator induced by $H_\mathrm{int}(t)$. The thermal bath state is now replaced by the non-thermal state
\begin{equation}
  \rho_\mathrm{B}^\text{non-th}=\UB\rho_\mathrm{B}^\mathrm{th}\UB^\dagger,
\end{equation}
so that equation~\eqref{eq_app_conditions_me1} now reads
\begin{multline}\label{eq_app_conditions_me1_replace_bath}
  \rho_\mathrm{WF}(t)=\Tr_\mathrm{B}\Big[U(t,0;H_\mathrm{int})(\mathds{1}_\mathrm{WF}\otimes\UB^\dagger)\\\times\left(\rho_\mathrm{WF}(0)\otimes\rho_\mathrm{B}^\mathrm{th}\right)(\mathds{1}_\mathrm{WF}\otimes\UB)U^\dagger(t,0;H_\mathrm{int})\Big].
\end{multline}
Owing to the cyclic property of the partial trace over the bath, i.e.,
\begin{equation}
  \Tr_\mathrm{B}\left(A(\mathds{1}_\mathrm{WF}\otimes\UB^\dagger)\right)\equiv\Tr_\mathrm{B}\left((\mathds{1}_\mathrm{WF}\otimes\UB^\dagger)A\right)
\end{equation}
for any $A$, and since $\UB^\dagger\UB=\mathds{1}_\mathrm{B}$, we can rewrite equation~\eqref{eq_app_conditions_me1_replace_bath} as 
\begin{multline}\label{eq_app_conditions_me2}
  \rho_\mathrm{WF}(t)=\Tr_\mathrm{B}\Big[(\mathds{1}_\mathrm{WF}\otimes\UB)U(t,0;H_\mathrm{int})(\mathds{1}_\mathrm{WF}\otimes\UB^\dagger)\\\times\left(\rho_\mathrm{WF}(0)\otimes\rho_\mathrm{B}^\mathrm{th}\right)\\\times(\mathds{1}_\mathrm{WF}\otimes\UB)U^\dagger(t,0;H_\mathrm{int})(\mathds{1}_\mathrm{WF}\otimes\UB^\dagger)\Big].
\end{multline}
Both equations~\eqref{eq_app_conditions_me1} and~\eqref{eq_app_conditions_me2} involve the \emph{same} thermal bath state, the only possible difference being due to the transformation of the time evolution operator and thus of $H_\mathrm{int}(t)$.

\end{document}